\documentclass[pdftex]{article}
\usepackage{icrctc07}

\title{Study of the performance and capability of the new ultra-fast 2 GSample/s FADC 
data acquisition system of the MAGIC telescope}
\shorttitle{Performance of the new 2~GSample/s FADCs of the MAGIC telescope}
\authors{
 D.~Tescaro$^{1}$,
 H.~Bartko$^{2}$,
 N.~Galante$^{2}$,
 F.~Goebel$^{2}$,
 T.~Jogler$^{2}$,
 R.~Mirzoyan$^{2}$,
 A.~Moralejo$^{1}$,
 T.~Schweizer$^{2}$,
 M.~Shayduk$^{3,}$$^{2}$,
 M.~Teshima$^{2}$\\
 on behalf of the MAGIC collaboration.}
\shortauthors{D.~Tescaro and et al.}
\afiliations{
 $^1$Institut de F\'\i sica d'Altes Energies, Edifici Cn., E-08193 Bellaterra (Barcelona), Spain\\
 $^2$Max-Planck-Institut f\"ur Physik, D-80805 M\"unchen, Germany\\
 $^3$Humboldt-Universit\"at zu Berlin, D-12489 Berlin, Germany
 }
\email{diegot@ifae.es}

\abstract{In February 2007 the MAGIC Air Cherenkov Telescope for gamma-ray astronomy was fully
upgraded with an ultra fast 2~GSamples/s digitization system.
Since the Cherenkov light flashes are very short, a fast readout can minimize the influence of the background from the light of the night sky.
Also, the time structure of the event is an additional parameter to reduce the background from unwanted hadronic showers. An overview of the performance of the new system and its impact on the sensitivity of the MAGIC instrument will be presented.
}

\begin{document}
\maketitle

\section{Introduction}

MAGIC is an Atmospheric Cherenkov Telescope situated on the Canary island of La Palma ($28.75^\circ$N,
$17.86^\circ$W, 2225~m a.s.l.).
Fundamental parameters of the telescope are a 17~m~(\O) mirror of parabolic shape (which preserves the time structure of the Cherenkov light flash), 
and a hexagonally shaped camera of 576 hemispherical  photo-multiplier tubes (PMT). 
The field of view of the camera is $\approx3.5^\circ$~\O.
The fast PMT analog signals are routed via optical fibers to the DAQ-sytem electronics where 
the signals are digitized and saved on disk. 
Further details can be found in
\cite{2005ICRC....5..359C}.

Until the end of January 2007 the signals were  digitized by a 300 MHz FADCs system.
On February 2007 the data acquisition was upgraded with ultra-fast FADCs capable to  digitize 
at the ultra-high speed of 2~GSample/s \cite{2007ICRCGoebel}.
In the old system, an additional stretching of the pulses was necessary in order to ensure a proper sampling of the signal.
After the upgrade, the stretching is no longer necessary and the width of the pulses is now $\simeq$~2.3~ns FWHM, half of the former value.
The new system enhances the telescope performance basically for two reasons: a reduction in the amount of NSB (Night Sky Background) light integrated with the real signal (due to a smaller integration window), and the possibility to reconstruct with a good time resolution the timing characteristics of the showers.
%
 

\section{Signal extraction and Image~Cleaning}

Currently a cubic spline is applied to find the maximum pulse within the useful range of the FADCs ($\sim$30~ns), and the position of the maximum of the spline is taken as arrival time of the signal at that pixel.
The intensity of the signal is then obtained by integrating the spline in a range of 7.5~ns.
The digital~filter method, used for the signal extraction for the analysis of MAGIC data in the past, has not yet been tuned to the characteristics of the new hardware.
The time resolution of each pixel has been estimated to be around 0.4~ns~RMS for a 40~phe signal through the study of the pixel time spread in calibration events, although this value may improve with the use of a more sophisticated pulse reconstruction.


\begin{figure*}[th]
\begin{center}
\includegraphics[width=0.95\textwidth,height=0.34\linewidth]{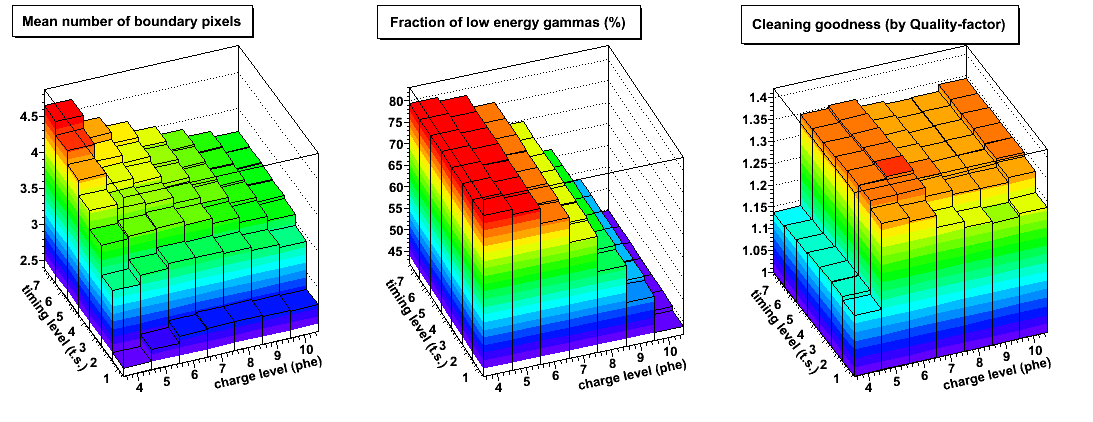}
\end{center}
\vspace{-0.5cm}
\caption{Cleaning features on low energy $\gamma$-MC (with a power law spectrum of index -2.6) depending on charge threshold ($x$ in phe) and time constrain ($y$ in time slices: 1~t.s.~=~0.5~ns) levels set. } \label{fig:cleaning}
\end{figure*}

The pixels which belong to the shower image are selected from the whole camera picture by the so-called Image Cleaning (IC) algorithm.
The procedure consists in setting a threshold signal value to select the so called "core pixels" (default 10~phe) and a second threshold to select or reject the neighbors as boundary pixels (default 5~phe).     
The idea behind the use of a further time constrain (imposing a limit on the arrival time difference between pixels) is to lower the pixel threshold, in order to have images with higher pixel number. 
In other words, a pixel is considered a boundary pixel if its charge is above the boundary threshold, at least one of its neighbors is a core pixel and its arrival time is within $\pm \Delta$t of that of the core pixel. 
Requiring this time coincidence allows keeping more information from boundary pixels, avoiding to confuse NSB signal with real image tails (since Cherenkov pulses are very short in time, the probability to include a fluctuation is very small).

IC levels have been optimized in this study.
Plots in figure \ref{fig:cleaning} represent the characteristics of different levels of cleaning. 
Only low energy events have been selected, through an upper cut in the signal of the two highest pixels (SIZE-2, a parameter which will not change by varying the tail cuts), since the main goal of the reduction of the tail cuts is to lower the energy threshold of the gamma event sample after image cleaning.
The entry on the $x$ axis (from 4~to~10) is the core pixel threshold imposed.
The threshold for  boundary pixels (not written) is always set as half of the core one.
On the $y$ axis (from 1~to~7) we have the second important setting level: the time coincidence window imposed between core and  boundary pixels (this is expressed in digitization time slices where 1~t.s.~=~0.5~ns).    

The first plot of figure \ref{fig:cleaning} represents the mean number of  boundary pixels in the images.
This number increases fast by lowering the charge threshold level
but a tighter time coincidence constrain keeps the number of  boundary pixels under control.
The second histogram describes the fraction of the triggered low energy gammas surviving the IC (at least two neighbor core pixels and three pixels in total are required).
More events means essentially a lower energy threshold (more low energy events have been kept).
The third plot is meant to classify the goodness of a certain cleaning configuration. The plotted value is the naive quality factor Q (enhancement of the signal to noise ratio) of the best cut in the ALPHA parameter.
For a point-like $\gamma$-ray source, an upper cut in this parameter rejects a good fraction of the isotropic background while keeping most of the signal.
In most of the explored parameter range the quality of ALPHA does not change significantly.
The sudden worsening of the image orientation reconstruction happening at a cleaning levels (4;2) phe is likely originated by the introduction of clusters of noise pixels (secondary islands).

\section{Time-related image parameters}

\begin{figure*}[t]
\begin{center}
\includegraphics[width=1.0\textwidth,height=0.25\linewidth]{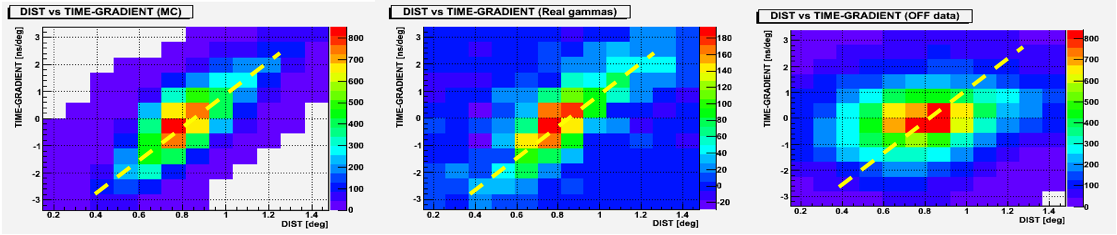}
\end{center}
\vspace{-0.7cm}
\caption{TIME GRADIENT - DIST correlation for MC- $\gamma$-ray (left), real- $\gamma$-ray (center) and OFF (right).} \label{fig:time_gradient}
\end{figure*}

In order to exploit the better time resolution of the new FADC system, new time-related image parameters have been introduced an characterized.
The TIME~RMS parameter estimate the arrival time spread of the Cherenkov photons in the pixels belonging the cleaned image ($\mu$,  $\gamma$-ray and hadron induced showers may have different characteristic time spread\cite{2006APh....25..342}).
Previous studies\cite{1999APh....11..363H} determined that along the major axis of gamma induced Cherenkov image is present a time structure. This can be well approximated as a linear TIME~GRADIENT.
The slope of this gradient is related to the angle between telescope and shower axes and moreover, for showers of a given direction, is well correlated with the Impact Parameter (IP) of the shower.

The possibility of using timing (multiplied by $c$) as third space coordinate of the charge distribution of the shower is at the moment still under investigation.

Monte~Carlo (MC) reliability is of extreme importance to perform a good gamma/hadron separation and 
energy reconstruction.
Both tasks are in fact  done comparing simulated  $\gamma$-ray to real data events.
The $\gamma$-ray-MC database has been updated accordingly to the new digitization features (for example the value of the time jitter of the PMTs used in the simulation has been tuned using the calibration events). 
Concerning the standard Hillas\cite{1985ICRC....3..445H} parameters the agreement between simulation and real events was known to be good\cite{2005ICRC....5..203M}, but also a comparison regarding the timing features is needed to ensure the goodness of the MC used. 

\begin{figure}[h!]
\begin{center}
\includegraphics[width=1.0\linewidth]{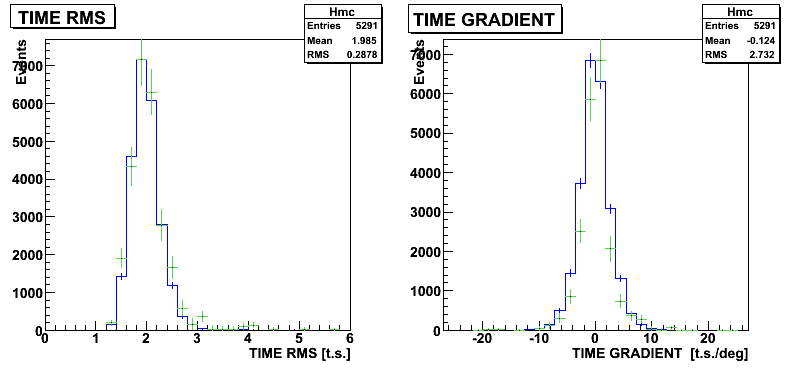}
\end{center}
\vspace{-0.7cm}
\caption{Time-related image parameters, real data-MC comparison. A cut greater than 150~phe in the SIZE parameter has been applied.
} \label{fig:time_parameters}
\end{figure}

A data sample of 5.6 h of Crab Nebula observation (8th 10th 16th and 18th of February 2007; zenith angle between $5^\circ$ and $30^\circ$) performed in wobble mode\footnote{Source is not pointed directly but $0.4^\circ$ away, in order to have simultaneous ON and OFF data \cite{1994APh.....2..137F}.} 
have been used for this purpose.
The distribution for the real data gammas is obtained as subtraction between ON and OFF distribution of the parameter considered. A very loose cut in the  $\gamma$/h separator parameter\footnote{The Random Forest (RF) \cite{2004NIMPA.516..511B} $\gamma$/h separation method provide a single parameter to estimate the probability of a shower to be  $\gamma$-ray or hadron initiated.} 
HADRONNESS ($<$0.8) was applied to preliminarly clean the distributions from clear hadron candidates (while removing nearly no gammas) events. 
The agreement turns out to be good (see figure \ref{fig:time_parameters}).

A proof that the TIME~GRADIENT - DIST correlation is present also in the real data  $\gamma$-rays and not only in the MC is given in figure \ref{fig:time_gradient}.
The left plot of the figure is done with pure  $\gamma$-ray-MC events while the plot on the center is the difference between ON and OFF distribution in the two-dimensional space considered. A correlation TIME~GRADIENT - DIST is present in both cases whereas the OFF data shows no correlation (plot on the right).

\section{Results and Conclusions}
\begin{figure*}[t]
\begin{center}
\includegraphics[width=0.84\textwidth,height=0.23\linewidth]{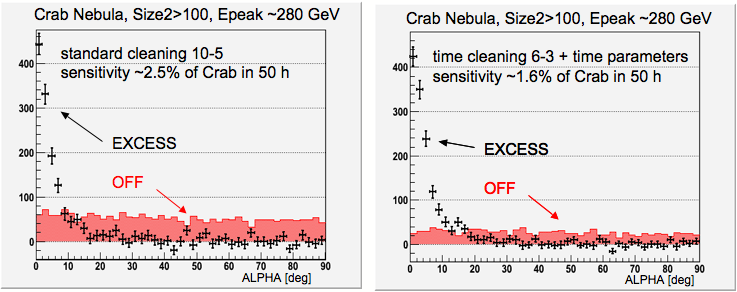}
\end{center}
\vspace{-0.5cm}
\caption{Comparison of two $\alpha$-plots between analysis 2 (left) and 3 (right). Keeping the same number of excess events background is reduced by about 50\%. Energy threshold: 280~GeV.} \label{fig:alpha}
\end{figure*}

Three different analyses of the same previously described Crab~Nebula data have been performed:
\begin{enumerate}
\item Using 10-5 IC and standard image parameters for $\gamma$/h separation (reference analysis).
\item Using an IC of 6-3 with also the time constrain (level 3~t.s.). The same standard parameters of analysis 1 for  $\gamma$/h separation.
\item Using the same time IC of analysis 2 and in addition to the standard parameters the TIME~RMS and TIME~GRADIENT were used for $\gamma$/h separation. 
\end{enumerate}

\begin{figure}[!b]
\begin{center}
\includegraphics[width=1.0\linewidth,height=0.48\linewidth]{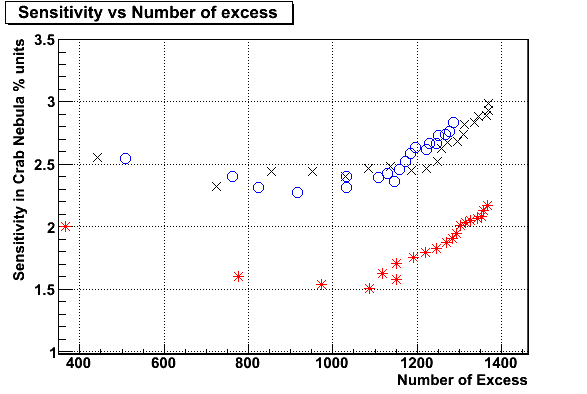}
\end{center}
\vspace{-0.7cm}
\caption{Sensitivity vs number of excess events ($E_{peak}$=280~GeV).
Cross circle and star points are refereed to analysis one two and three respectively.
} \label{fig:sigma}
\end{figure}

In figure \ref{fig:alpha} we compare two $\alpha$-plots obtained from analysis 2 and 3.
Time parameters allows $\sim$50\% better background suppression keeping the same amount of excess events. Similar improvements is seen at all energies.\\
In figure \ref{fig:sigma} is shown a more general comparison:
The telescope sensitivity\footnote{Defined as the flux that gives an excess equal to five time the square root of the background.} as function of the number of excess events in the signal region for different HADRONNESS cuts.
The ALPHA was optimized for each point independently.\\
%
Improvements (of ~15\%) have been found also in the event energy reconstruction. In fact the TIME GRADIENT gives information about the real IP of the shower and therefore it helps to distinguish distant high energy showers from closer, low energy ones.\\
We concentrated here on the $\alpha$-analysis approach but further studies using the $\theta^2$ method are already planned. How to apply this method in a source position independent framework and how timing could be use for the optimization of DISP are the first issues scheduled. 

%
\section{Acknowledgements}
We want to thank the IAC (Instituto Astrofisico de Canarias), the German BMBF and MPG, the Italian INFN, the Spanish CICYT, the Swiss ETH and the Polish MNiI.
\nocite{UltrafastFADC-NS}
\nocite{2001ICRC....7.2845M}

\bibliography{icrc0662}
\bibliographystyle{abbrv}

\end{document}